\documentclass[12pt,preprint]{aastex}

\newcommand{\grs}      {\mbox{\rm\,GRS~1758--258}}

\newcommand{\onee}     {\mbox{\rm\,1E~1740.7--2942}}
\newcommand{\cyg}      {\mbox{\rm\,Cyg~X--1}}
\newcommand{\her}      {\mbox{\rm\,Her~X--1}}
\newcommand{\cygtwo}      {\mbox{\rm\,Cyg~X--2}}
\newcommand{\ssfour}      {\mbox{\rm\,SS~433}}
\newcommand{\lmcx}      {\mbox{\rm\,LMC~X--3}}
\newcommand{\lmcfour}      {\mbox{\rm\,LMC~X--4}}

\begin{document}

\title{Orbital and Super-Orbital Periods of \onee\ and \grs}

\author{D. M. Smith\altaffilmark{1},
W. A. Heindl\altaffilmark{2}, J. H. Swank\altaffilmark{3}}

\altaffiltext{1}{Space Sciences Laboratory, University of California Berkeley, 
Berkeley, CA 94720}
\altaffiltext{2}{Center for Astrophysics and Space Sciences, Code 0424, University
of California San Diego, La Jolla, CA 92093}
\altaffiltext{3}{NASA Goddard Space Flight Center, Code 666, Greenbelt, 
MD 20771}

\begin{abstract}

Five years of \it Rossi X-ray Timing Explorer (RXTE) \rm observations
of the Galactic black-hole candidates \onee\ and \grs\ show a periodic
modulation with amplitude 3-4\% in each source at $12.73 \pm 0.05$~dy
and $18.45 \pm 0.10$~dy, respectively.  We interpret the modulations
as orbital, suggesting that the objects have red-giant companions.
Combining the \it RXTE \rm data with earlier data \citep{Zh97} from
the Burst and Transient Source Experiment on the \it Compton Gamma-Ray
Observatory\rm, we find a long period or quasi-period of about 600~dy
in \onee, and a suggestion of a similar 600-dy period in \grs.  These
timescales are longer than any yet found for either precessing systems
like \her\ and \ssfour\ or binaries like \lmcx\ and \cyg\ with more
irregular long periods.

\end{abstract}


\section{Introduction}

The Galactic-bulge x-ray sources \onee\ and \grs\ are generally called
black-hole candidates due to the similarity of their x-ray spectral
and timing behavior to that of \cyg\ in its usual hard state.  Like
\cyg, both sources occasionally enter an intermediate or soft state,
but the evolution of their spectral hardness and luminosity is very
different \citep{Sm02}.  Both have prominent, bright radio lobes about
an arcminute in size \citep{Mi92,Ro92}, while \cyg\ does not
\citep{Ma96}, showing only a milliarcsecond jet near its core
\citep{St01}.

The counterparts of \onee\ and \grs\ are unknown due to high
extinction, therefore there are no orbital solutions or estimated
masses.  For \grs, \citet{He02} recently used \it Chandra \rm
data to confirm the association of the x-ray source with the radio
core source ("VLA-C") and \citet{Cu01} did the same for \onee.
Marti et al. (1998) identified two candidate counterparts
to VLA-C/\grs\ in \it I- \rm and \it K-\rm band images. The brighter
and closer candidate was found through multi-band photometry and
near-infrared spectroscopy to be a likely K0 III giant.  Revised
astrometry \citep{Ro02} of infrared observations by \citet{Ei01a}
confirm that this star (``star A'') is consistent with VLA-C at the
3$\sigma$ level. \citet{Ma00} and \citet{Ei01a} agree on several
possible high-mass candidates for the companion of \onee\ in its more crowded
and obscured field, but a low-mass companion would
be unobservable at the current \it K\rm-band sensitivity.

\section{Observations}

We use a five-year (1997--2001) series of observations by the \it Rossi
X-ray Timing Explorer (RXTE) \rm of \onee\ and \grs\
\citep{Ma99,Sm01,Sm02}.  The observations, each of 1000--1500 s, were
taken approximately weekly in 1997--2000 and twice weekly since early
2001.  Monthly observations in 1996 are too sparse to improve
our results and are not included.  Data cannot be taken from late November
to late January of each year when the Sun is close to the Galactic
Center.

We use the Proportional Counter Array (PCA), layer 1, in the range 2.5
to 25~keV.  Instrumental background has been subtracted using the
``faint source'' model and Galactic diffuse emission has been
subtracted using pointings to nearby fields without bright point
sources. To compensate for gain changes, we accumulate counts in bands
of constant energy, not channel number.  We offset-point by about a
half degree from each source to avoid nearby bright sources.  Details
of the offsets and background pointings are given in \citet{Ma99}.

Since our orbital signals are small, we made two checks for systematic
errors.  Different observations use different subsets of the 5
detectors of the PCA.  We repeated the entire analysis below using
only the third detector, which is always on, and found no change
except for the expected statistical degradation.  Slight changes in
the PCA field of view due to changes in the roll angle of the
spacecraft take place over months, and so cannot mimic the shorter
orbital periods.

\begin{figure}
\epsscale{0.42}
\plotone{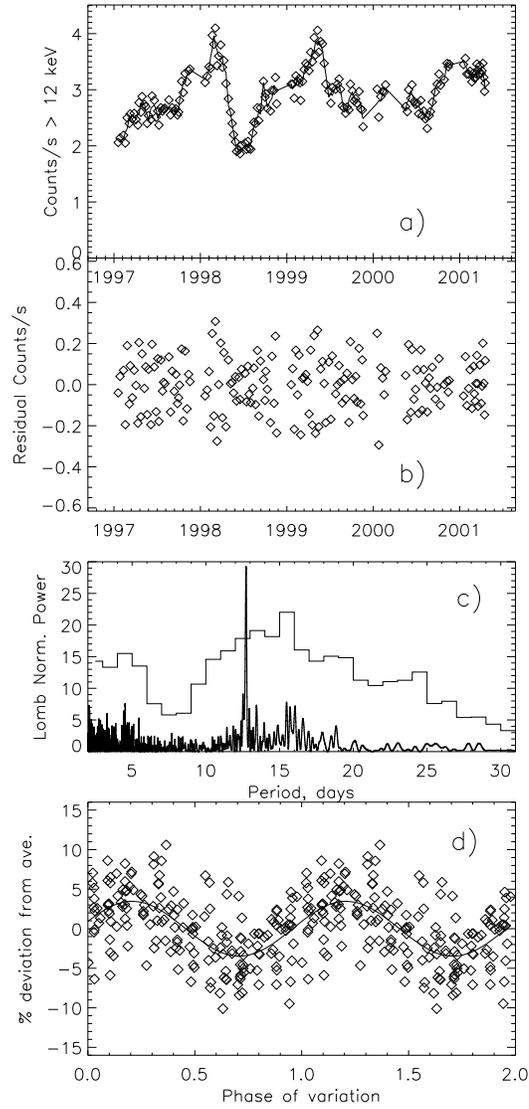}
\caption{
Analysis of the lightcurve of
\onee.  a): Count rate $>$ 12~keV per detector
of the PCA.  The error bars are smaller than the
plotting symbols.  The smoothed curve is described in $\S$ 3.
b):  The same data high-pass filtered by
subtraction of the smoothed curve. c): Fine histogram: Lomb-Scargle
periodogram of the data in panel b).  Coarse histogram: highest
values achieved by applying the same analysis to scrambled data
(see text). d): Data from panel b) folded on the period of the peak in
panel c), and normalized to the average count rate.  The smooth curve
is the best fit sine function.}
\end{figure}

\begin{figure}
\epsscale{0.42}
\plotone{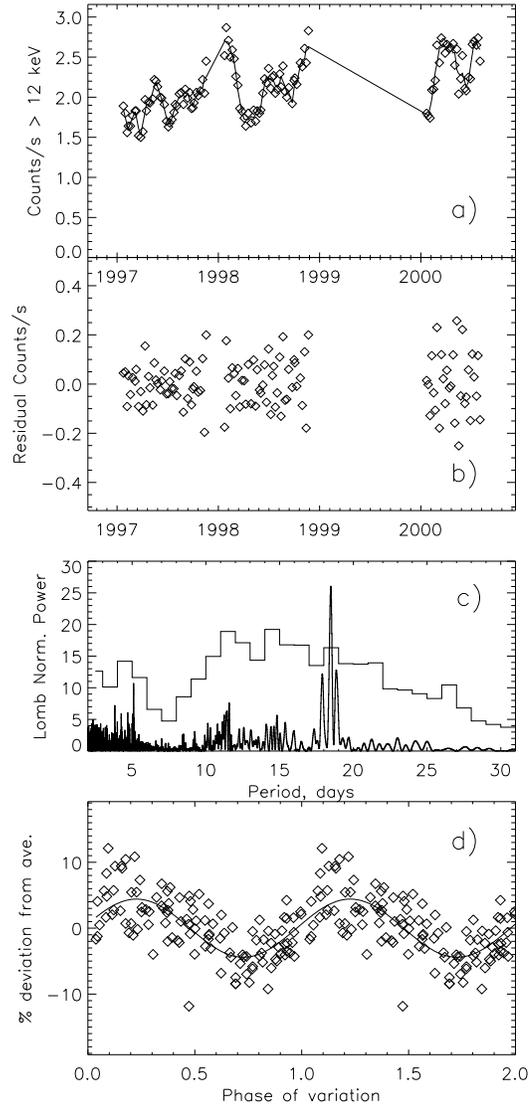}
\caption{Same as Figure 1 for \grs.  \grs\ spent more of its time
in the soft state, so more data have been removed.}
\end{figure}

\section{Analysis and results}

\subsection{Orbital modulation}

During long intervals of relatively stable hard
state emission, these sources show low-amplitude modulations that are
most naturally interpreted as orbital.  Figures 1a and 2a
show the 1997--2001 background-subtracted lightcurves
from 12--25~keV.  There are gaps due both to the annual solar
constraint and to transitions to the intermediate and soft states
\citep{Sm01,Sm02}.  Luminosity variations are so large during the
state transitions and soft periods that they decrease the significance
of the orbital measurements.  The data cut is impartial in that only
the spectral power-law photon index is used to determine which parts
of the data to remove: it is harder than 2.0 in the surviving data.

We high-pass-filtered the data by subtracting out a smoothed version
of the lightcurve (Figures 1a and 2a) leaving only high-frequency
residuals (Figures 1b and 2b).  To make the smoothed curve, we
replaced each data point with a value generated by fitting a
polynomial of order $P$ to all data within $N$ days of it.  Raising
$N$ or lowering $P$ increases the amount of smoothing.  The figure
shown is for $N=10$ and $P=1$, but the results are not highly
sensitive to these values.

We then took a Lomb-Scargle periodogram of the residuals with the
result shown in Figures 1c and 2c: peaks at
$12.73 \pm 0.05$~dy in \onee\ and $18.45 \pm 0.10$~dy in \grs.  The
errors are the half width at half maximum of the peaks.  Although
these values would be near the Nyquist frequency for weekly
sampling, the irregularity in the sampling times extends the useful
frequency range many times higher.

Figures 1d and 2d show the residuals folded on the best-fit periods
and divided by the average count rate.  The phases are referenced to
0h UT on 1 January 1996.  The best-fit amplitudes are $(3.43 \pm
0.26)\%$ for \onee\ and $(4.42 \pm 0.32)\%$ for \grs. 

\begin{figure}
\epsscale{0.75}
\plotone{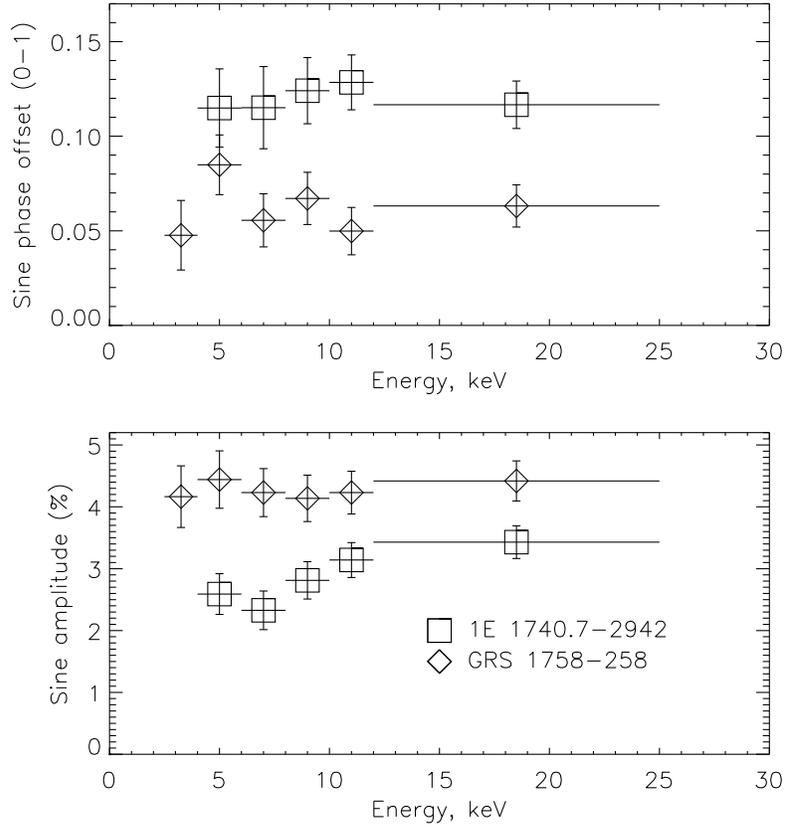}
\caption{Dependence of the parameters of the best-fit orbital sinusoids
on energy.  Top: phase (at 0h UT on 1 January 1996).  Bottom: amplitude.}
\end{figure}

When the data from each source are divided into three roughly equal
intervals, the orbital modulation is seen in all three as the highest
periodogram peak in the range 1--30~dy.  We also
repeated the analysis in 2.5--4.0~keV, 4.0--6.0~keV, 6.0--8.0~keV,
8.0--10.0~keV, and 10.0--12.0~keV bands.  The modulations are
not as strong, but in every band they are the
highest peaks from 1--30~dy.  Neither the amplitude nor phase
varies significantly with energy, although amplitude may increase
slightly with energy in \onee\ (Figure 3). The 2.5--4.0~keV band is not
used for \onee\ due to its high absorption column.

We demonstrated by simulations that our process neither creates
spurious signals nor destroys real ones.  To look for spurious
signals, we randomly permuted the values of each triplet of data
points in Figures 1a and 2a.  This creates data with the same sampling
times, average, and long-term evolution as the real set but with any
high-frequency signals destroyed.  A thousand differently-permuted
datasets were analyzed for each source, giving the coarse histograms
in Figures 1c and 2c, which show the highest power seen in one-day-wide
period bins in any of the 1000 trials.  It is clear
that peaks the size of the real signals do not occur.

To test the usefulness of the high-pass filtering, we added to the
original data small sinusoidal signals with 3\% to 20\% of the mean
flux and 10--20~dy periods.  We then did the analysis with and without
filtering.  We found that filtering improves the sensitivity to
artificial signals by a factor of 2 to 4, varying with the period and
data set.  Artificial signals with the amplitude of the real
ones can be seen, but only by filtering first, proving the necessity
of the procedure.

\subsection{Super-orbital period modulation}

\begin{figure}
\epsscale{0.75}
\plotone{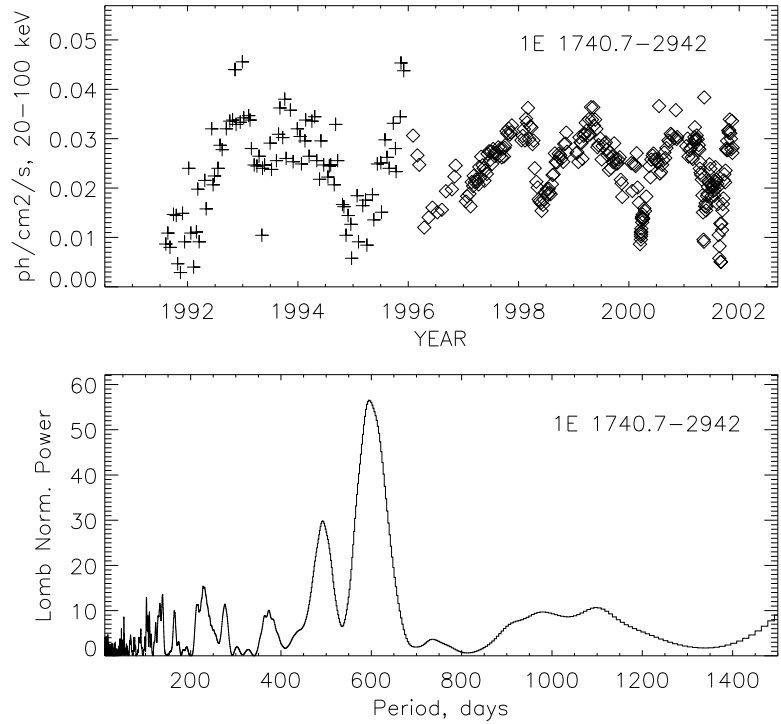}
\caption{BATSE fluxes for \onee\ from \citet{Zh97} (crosses) and
PCA data extrapolated to the same energy range (diamonds).  
Top: flux vs. time, 20--100~keV. Bottom: Lomb-Scargle 
periodogram.}
\end{figure}

\begin{figure}
\epsscale{0.75}
\plotone{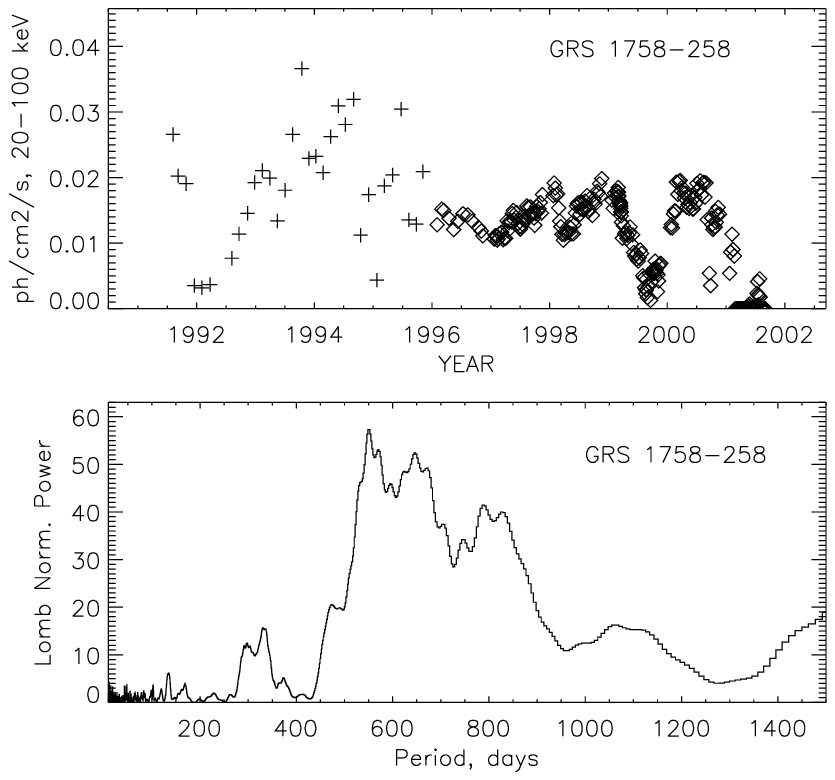}
\caption{Joint BATSE/PCA lightcurve and periodogram for \grs.}
\end{figure}

To search for longer periods, we combined data \citep{Zh97} from the
Burst and Transient Source Experiment (BATSE) on the \it Compton
Gamma-Ray Observatory \rm with our \it RXTE\rm/PCA data, extrapolating
the PCA data to the BATSE range of 20--100~keV.  For most black holes
in the hard state this is not a bad approximation: only beyond 100~keV
do the spectra steepen significantly.  Figures 4 and 5 show the
combined lightcurves and their periodograms.  The major
dips in the lightcurves all correspond to softenings of the spectra.

The periodogram for \onee\ shows a clear peak around 600~dy.  Each
individual 600-dy cycle is visible if the single low point from BATSE
in 1993 is included.  The data for \grs\ are less conclusive.  The
BATSE data are poorer for \grs, due in part to its slightly lower
luminosity but probably also to the proximity of the bright Z-source
GX~5--1.

\section{Interpretation}

If the 12- and 18-dy modulations are orbital, then the
companions of \onee\ and \grs\ must be giant stars to fill their Roche
lobes.  For \grs, infrared observations rule out a high-mass companion
\citep{Ma98,Ei01a} and therefore make wind accretion less likely.  Star A of
\citet{Ma98}, a K0 III giant, would almost exactly fill its Roche lobe
with a 10M$_{\odot}$ companion and an 18-dy orbit \citep{Sm00,Ro02}.
Perhaps both systems have companions on the first giant branch and are
in a long-lived evolutionary phase with accretion driven by the
gradual nuclear evolution and expansion of the secondaries, as
has been suggested for long-period neutron-star binaries like
\cygtwo\ \citep{We83}.  \citet{Ki97}, however, found that such
systems should have unstable disks and appear as transients
if the primary is a black hole. 

Low-amplitude ($\sim$7\%) x-ray modulation with a similar shape was
recently discovered in another persistently luminous black-hole
candidate, \lmcx, at its previously-known orbital period of 1.7~dy
\citep{Bo01}.  \cyg\ also shows low-amplitude orbital x-ray
modulation, which is not observed in the soft state.  \citet{We99}
offered two interpretations: partial absorption by the optically thin
wind of the secondary \citep[see also][]{Br99}, and partial
obscuration of an extended emission region (corona) by the accretion
stream.  The lack of modulation in the soft state was explained in the
first picture by increased ionization and decreased mass in the wind,
and in the second picture by shrinkage of the corona.

Because their amplitude and phase are independent of energy (Figure
3), the modulations of \onee\ and \grs\ are consistent with an
extended emission region and an optically thick absorber such as the
companion star.  Wind absorption would produce more modulation at
lower energy, as is seen in \cyg\ \citep{Br99}.  Assuming the
secondary itself is the occulter, then given the size and inclination
of the binaries we could determine the size of the hard-state emission
region from the amplitude of the modulation.  We are planning infrared
observations of star A over the orbit to determine the primary mass,
system size, and inclination using radial velocity, velocity
broadening, and ellipsoidal modulation measurements.

The prominent radio lobes of \onee\ and \grs, appearing far from
the core sources, imply strong collimation of the jets.
\citet{Sp97} show that in a jet collimated by the
poloidal magnetic field of an accretion disk, collimation is
proportional to the ratio of the outer to inner disk radii.  The
very large disks implied by the long binary periods in these
systems may support this model.

The 600-dy super-orbital periods we report here are the longest seen
in any system so far.  \her\ and \lmcfour\ have very regular long
x-ray periods, about 35~dy and 30~dy respectively \citep{Ta72,La81}, which
are explained by disk precession with periodic partial
obscuration of the x-ray emitting regions \citep[e.g.][]{Wi99}.  Jet
precession at a period of about 162~dy is observed via Doppler shifts
of emission lines in \ssfour\ \citep{Ma79,Ei01b}, and is presumed to
be caused by precession of the disk.  Other systems, such as the
black-hole candidate \lmcx, show more irregular variations, associated
with spectral state changes that are not consistent with disk
obscuration \citep{Pa00,Wi01}.  \citet{Br01} review recent models of
long-period variability involving changes in accretion rate either due
to a limit cycle caused by x-ray evaporation of the outer parts of the
accretion disk \citep{Sh86} or changes in the companion which cause it
to fill its Roche lobe only occasionally \citep{Wu01}.  \cyg\ in
different epochs has been reported to show long periods at 294~dy and
142~dy \citep{Pr83,Br99}.

\citet{Og01} calculated the susceptibility of disks in x-ray binaries
to radiation-driven warping.  We find the binary separations of \onee\
and \grs\ to be about (2.5--5)$\times 10^{12}$ cm for black hole
masses of 3--20 M$_\sun$.  Using Figure 7 of \citet{Og01}, we find that
for a secondary of 1.1 M$_\sun$ (i.e. star A), black hole masses
greater than 6 M$_\sun$ fall in the ``indeterminate instability
zone'', with possible cycling between flat and warped disks.  Such
cycling might explain long, semi-regular periods with spectral
changes, even with constant mass transfer from the companion.  For
lower masses of the primary, the systems would fall in the regime of
persistent warping and stable precession, as in \ssfour.

A recent observation of \grs, however, suggests that variations in
accretion do play a role in its long-term variability.  An extreme
luminosity drop is visible at about 2002.25 in Figure 5.  \citet{Sm01}
found that \grs\ entered a soft state suddenly and then decayed
exponentially with a one-month timescale.  Exponentially decaying
emission in the soft state is easily explained by the steady draining
of a thin accretion disk with little or no mass input.

This work was supported by NASA grant NAG5-7265.

\end{document}